\documentclass[aps,pra,twocolumn,showpacs,showkeys,10pt]{revtex4-1}

\usepackage{amsmath,amssymb,upgreek,bbm}
\usepackage{graphicx,color}

\graphicspath{ {.} {./pics/} }

\newcommand{\rme}{\mathrm e}
\newcommand{\rmi}{\mathrm i}
\newcommand{\rmd}{\mathrm d}

\renewcommand{\Re}{\mathop{{\rm Re}}}

\begin{document}

\title{Generation, dynamical buildup and detection of bi- and mulipartite entangled states in cavity systems}

\author{D. Pagel}
\email{pagel@physik.uni-greifswald.de}
\author{H. Fehske}
\affiliation{Institut f\"ur Physik, Ernst-Moritz-Arndt-Universit\"at, 17487 Greifswald, Germany}

\begin{abstract}
We inspect different quantum optical setups from the viewpoint of entanglement generation and detection.
As a first step we consider a planar semiconductor microcavity and optimize the Bell-type correlations and their robustness against dephasing to create strong bipartite entanglement between polariton branches, which subsequently can be transfered to the emitted photons.
In a second step, in order to create multipartite entangled light, we place the microcavity in an optical resonator driven by pump pulses with a frequency comb spectrum.
For this system we show how phase matching of all comb modes can be achieved and will lead to indistinguishable scattering processes causing entanglement among every mode.
Finally we demonstrate the buildup of entanglement in the dissipative dynamics of emitters coupled to a single cavity photon mode driven by an external laser.
From a Floquet master equation approach we find that entanglement production predominates during the first few laser oscillation periods.
\end{abstract}

\pacs{
  03.65.Yz,     
  03.67.Bg,     
  42.50.-p,     
}

\keywords{Quantum Optics, Open Systems, Entanglement Production}

\maketitle

\section{Introduction}
Quantum entanglement~\cite{sch35, epr35, hhhh09, nc10}, embodying a certain kind of correlation between subsystems without classical counterpart~\cite{wer89}, is the fundamental resource for future quantum computation and communication applications~\cite{fey82, bw92, llo96}.
The most elementary realizations are Bell states in bipartite situations~\cite{bel64} and GHZ or W states~\cite{ghz89, dvc00} in the event that more than two subsystems are entangled.
In the multipartite case, a state is called partially entangled if some (but not all) subsystems can be separated, and fully entangled if subsystems can not be separated.

Theoretically, entanglement can be determined by so-called witnesses, exhibiting negative expectation values for entangled states~\cite{hhh96, hhh01}.
A witness is an observable and thus can be measured in experiments.
The problem is, that a positive expectation value does not allow to decide whether or not the state is entangled.
To overcome this drawback an optimization was introduced~\cite{lkch00}, yielding necessary and sufficient conditions for the detection of entanglement~\cite{sv09a, sv11a}.
In order to quantify the amount of entanglement of a given state, one has to construct a proper entanglement measure~\cite{vprk97, vidal00, bra05}.
Albeit this has been done for bipartite systems~\cite{bra05, sv11a, sv11b}, the general multipartite case resists a complete solution so far~\cite{sv13, sw16}.

Also from an experimental point of view, despite recent progress~\cite{vsbysc01, cz04, vptf06, marcfft14}, the generation and control of entangled states is a still challenging task, because the necessary quantum coherence usually rapidly decays.
Entanglement distillation of non-maximally entangled states may be a route to overcome this problem~\cite{bbpssw96, bbpssw97, hhh97}.
Alternatively, the experimental setup can be optimized to generate a high amount of entanglement.
In this respect, optical realizations are promising~\cite{dgshpes11}.
Entanglement generation then relies on strong interactions in hybrid light-matter systems.
This regime is accessible in cavity-quantum or circuit-quantum electrodynamics setups~\cite{ndhmhsgrzhsm10, aga13, yfakss17} and is usually described by the famous Dicke model of two-level emitters coupled to a cavity photon mode~\cite{dic54}.
Other approaches are based on parametric down-conversion in nonlinear crystals~\cite{kmwzss95, ak04, gsvcrtf15}, biexciton decay in quantum dots~\cite{bspy00, hsk03}, parametric interactions of cavity photons and semiconductor excitons in two-dimensional microcavities~\cite{wnia92, hwsopi94, lang04, cbc05}, or nitrogen-vacancy centers in photonic crystals~\cite{yyxfo11, wkkb14}.

In this paper, we address the generation, evolution and detection of entanglement for three different quantum optical settings.
First, we take up the idea that strong bipartite entanglement can be created by a two-dimensional semiconductor microcavity driven by a pump-pulse train~\cite{pfsv12}.
Tuning the system parameters, Bell-type correlations can be enforced, transferred to nonclassical light, and verified by the Schmidt decomposition of the system's state.
In this way, we identify the optimal parameter region for the generation of highly entangled light.
Second, we propose a scenario where a planar microcavity is placed inside an optical resonator that is driven by lasers with a frequency comb spectrum.
We show, that this system is capable of distributing entanglement among each mode of the comb leading to a highly multipartite entangled (continuous variable) state which again can be proved in the emitted light.
Third, we demonstrate, analyze and quantify a dynamical buildup of (bipartite) entanglement among two emitters placed inside a laser-driven cavity.
We find that the entanglement during the first few modulation periods is much higher than the asymptotic value at later times.
This offers the possibility to enhance the entanglement by successive pump pulses.

The article is organized accordingly.
Section~\ref{sec:microcavity} introduces the driven microcavity model and gives the Schmidt number at fixed dephasing in dependence on the detuning and the polariton splitting to binding energy ratio.
Section~\ref{sec:comb} deals with multipartite entangled light starting from a frequency comb spectrum.
Specifically, the phase-matching polariton interactions were studied in Sec.~\ref{ssec:phasematch}, the ground state of the system is evaluated in Sec.~\ref{ssec:groundstate}, and its entanglement properties are analyzed in Sec.~\ref{ssec:entanglement}.
Section~\ref{sec:dyn} describes the dynamics of entanglement between two emitters strongly coupled to a laser-driven cavity in terms of the Dicke model.
Finally, Sec.~\ref{sec:conc} provides our conclusions.

\section{\label{sec:microcavity}Planar semiconductor microcavity}
\begin{figure}
  \includegraphics[width=\linewidth]{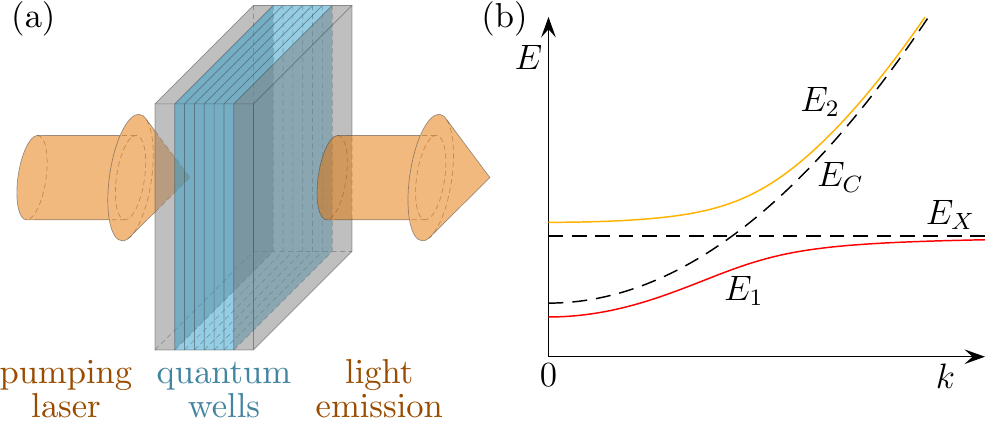}
  \caption{\label{fig:qwell}(a) Pumped planar semiconductor microcavity consisting of quantum wells between mirrors; (b) dispersions of semiconductor excitons ($E_X$), cavity photons ($E_C$) and polaritons ($E_{1,2}$).}
\end{figure}

We consider a planar semiconductor microcavity driven by external optical pumping [see Fig.~\ref{fig:qwell}(a)].
Such a system is usually described in terms of (bosonic) polaritons~\cite{usui60, ty99, csq01, cbc05}, which provides an efficient framework to investigate polariton parametric scattering in momentum space~\cite{ciu04, lang04}.
An alternative approach, which is based on the Heisenberg equations of motion for the semiconductor exciton and cavity photon operators, is the so-called dynamics controlled truncation formalism~\cite{as94, sg96, pdssrg08a, pdssrg08b}.

Polaritons are hybrid light-matter quasiparticles that arise from the strong coupling between semiconductor excitons and cavity photons~\cite{wnia92, hwsopi94}.
During the process of polariton formation two branches emerge, whereby an avoided crossing of exciton and photon modes takes place [see Fig.~\ref{fig:qwell}(b)].
The energies of these polariton branches are
\begin{equation}\label{disp}
  E_j(\mathbf{k}) = \frac{1}{2} \Big[E_C(\mathbf{k}) + E_X \pm \sqrt{(E_C(\mathbf{k}) - E_X)^2 + 4 \hbar^2 \Omega^2} \Big],
\end{equation}
where $j = 1,2$ is the branch index, $\mathbf{k} = (k_x, k_y)^T$ denotes the two-dimensional wave vector with modulus $k = |\mathbf{k}|$, $E_C(\mathbf{k}) = E_C \sqrt{1 + (\hbar c k / E_C)^2}$ is the cavity photon energy with $E_C(0) = E_C$, $E_X$ is the (dispersionless) exciton energy, and $2 \hbar \Omega$ gives the vacuum Rabi splitting.
The polariton dispersions in Fig.~\ref{fig:qwell}(b) show a clear asymmetry between the lower and upper branch which is a consequence of the normalized detuning $\delta = (E_C - E_X) / (2 \hbar \Omega)$ as well as of the cavity $E_C(\mathbf{k})$ and exciton $E_X$ energies.

Pumping with a laser resonantly injects polaritons with fixed wave vector and energy.
The Coulomb interaction between the exciton constituents (electrons and holes) in combination with saturation effects in the course of bosonization of their actual fermionic excitations yields an effective polariton pair interaction~\cite{ty99, oss98}.
Therefore, two pumped polaritons can scatter into pairs of signal and idler polaritons, if energy and momentum is conserved (phase-matching)~\cite{sbsswr00, sdssl05, pdsss09}.
One distinguishes single-pump from mixed-pump processes, where the two initial polaritons are injected by a single or two different pumped modes, respectively.
The indistinguishability of the scattering channels for specific pump configurations is the key for the generation of entangled polaritons~\cite{ciu04, ab12, pevwr14, evwp15}.
Since the polaritons have a large photon component any polariton entanglement can efficiently be transfered to the emitted photons~\cite{wnia92, csq01}.

If a pump laser is split into multiple beams that are aligned on a cone with incident angle below the magic angle~\cite{lang04, sbsswr00}, the generation of multipartite entangled W states becomes possible~\cite{pfsv13}.
This setup involves scattering within the lower polariton branch and seems therefore feasible in experiments~\cite{pf15}.
The multipartite correlations of the photons in the output channels can be identified with so-called multipartite entanglement witnesses which requires the solution of separability eigenvalue equations~\cite{sv13}.
Such a solution is given in Ref.~\cite{pfsv13} for a general witness constructed from W states allowing for the detection of light entanglement in the presence of lossy media.

Strong bipartite entangled states can be generated if the upper polariton branch is driven by a pump pulse train.
Assuming a train of $N$ individual modes, the state of the corresponding $N$ branch-entangled polariton pairs takes the form~\cite{pfsv12}
\begin{equation}
  | \psi \rangle = \prod_{n=1}^N \big( \alpha_n \hat{a}_{1, \text{s}n}^\dagger \hat{a}_{2, \text{i}n}^\dagger + \sqrt{1 - \alpha_n^2} \hat{a}_{2, \mathbf{k}_{\text{s}n}}^\dagger \hat{a}_{1, \mathbf{k}_{\text{i}n}}^\dagger \big) | \text{vac} \rangle \,.
\end{equation}
In this equation, $| \text{vac} \rangle$ is the polariton vacuum and $\hat{a}_{j, \text{x}n}^\dagger$ creates a signal ($\text{x} = \text{s}$) or idler ($\text{x} = \text{i}$) polariton in the lower ($j = 1$) or upper ($j = 2$) branch.
The index $n$ indicates that two pumped polaritons from mode $n$ of the train scatter into signal and idler, i.\,e., given a pump wave-vector $\mathbf{k}_{\text{p}n}$ we have $\mathbf{k}_{\text{s}n} = \mathbf{k}_{\text{p}n} + \mathbf{q}_n$ and $\mathbf{k}_{\text{i}n} = \mathbf{k}_{\text{p}n} - \mathbf{q}_n$ for a phase-matching scattering wave-vector $\mathbf{q}_n$.
The constants $\alpha_n$ characterize the properties of the material and explicitly read~\cite{pfsv12}
\begin{equation}
  \alpha_n = V_{\mathbf{k}_{\text{p}n}, \mathbf{k}_{\text{p}n}, \mathbf{q}_n}^{1, 2, 2, 2} \big[ \big( V_{\mathbf{k}_{\text{p}n}, \mathbf{k}_{\text{p}n}, \mathbf{q}_n}^{1, 2, 2, 2} \big)^2 + \big( V_{\mathbf{k}_{\text{p}n}, \mathbf{k}_{\text{p}n}, \mathbf{q}_n}^{2, 1, 2, 2} \big)^2 \big]^{-1/2} \,.
\end{equation}
Here, 
\begin{eqnarray}
  V_{\mathbf{k}, \mathbf{k}', \mathbf{q}}^{j_1 j_2 j_3 j_4} &=& 12 E_b M_{1 j_1 \mathbf{k} + \mathbf{q}} M_{1 j_2 \mathbf{k}' - \mathbf{q}} M_{1 j_3 \mathbf{k}} M_{1 j_4 \mathbf{k}'} \nonumber\\
    && -\frac{16 \pi}{7} \hbar \Omega \big( M_{2 j_1 \mathbf{k} + \mathbf{q}} M_{1 j_2 \mathbf{k}' - \mathbf{q}} M_{1 j_3 \mathbf{k}} M_{1 j_4 \mathbf{k}'} \nonumber\\
    && + M_{1 j_1 \mathbf{k} + \mathbf{q}} M_{1 j_2 \mathbf{k}' - \mathbf{q}} M_{1 j_3 \mathbf{k}} M_{2 j_4 \mathbf{k}'} \big)
\end{eqnarray}
is an effective branch-dependent potential with the matrix elements of the Hopfield transformation~\cite{hop58}
\begin{subequations}\begin{eqnarray}
  M_{1 1 \mathbf{k}} &=& M_{2 2 \mathbf{k}} = \frac{1}{\sqrt{1 + \rho_\mathbf{k}^2}} \,, \\
  M_{1 2 \mathbf{k}} &=& -M_{2 1 \mathbf{k}} = \sqrt{1 - M_{1 1 \mathbf{k}}^2} \,, \\
  \rho_\mathbf{k}    &=& \frac{E_2(\mathbf{k}) - E_C(\mathbf{k})}{\hbar \Omega} \,.
\end{eqnarray}\end{subequations}
and the exciton binding energy $E_b$.

The non-classical correlations of the emitted photons can be analyzed using the Schmidt number (SN)~\cite{th00, sbl01, sv11b}, which quantifies the entanglement of a bipartite state based on the superposition of product states.
For pure states the SN arises from the Schmidt decomposition of the state, that is, the SN counts the nonzero coefficients~\cite{nc10}.
Hence, every separable state has a SN of 1.
A maximally entangled Bell state has $\text{SN} = 2$.
$\text{SN} > 2$ is possible for subspace dimensions larger than one.
A generalization to mixed quantum states can be achieved by a convex roof construction.

\begin{figure}
  \includegraphics[width=0.7\linewidth]{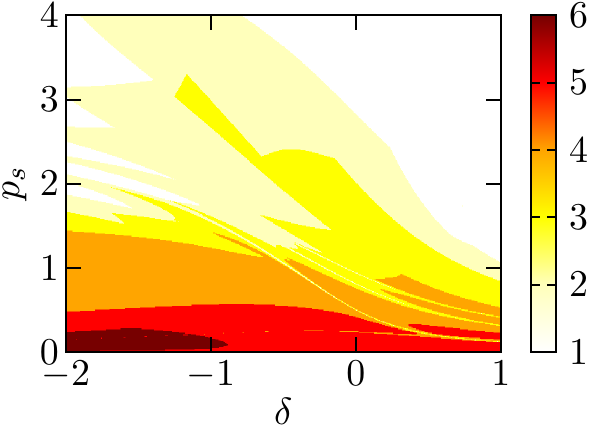}
  \caption{\label{fig:SN}Schmidt number for $N = 3$ in dependence on detuning $\delta$ and polariton-splitting to binding-energy ratio $p_s$ at fixed dephasing time $\Delta t = 300\,$fs.}
\end{figure}

Here, we use SN witnesses~\cite{bchhkls02, sv11a, pfsv12} to quantify the entanglement of the emitted photons that decreases due to propagation in dephasing channels, see Ref.~\cite{pfsv12} for details of this formalism.
Our goal is to maximize the resulting SN in dependence on the normalized detuning $\delta$ and the ratio of polariton splitting to exciton binding energy $p_s = 2 \hbar \Omega / E_b$ to find the best system parameters for experimental implementation.
In Fig.~\ref{fig:SN} the SN for $N = 3$ with upper bound $2^N = 8$, given for a fixed dephasing time $\Delta t = 300\,$fs, is obtained by numerical optimization of the SN witness and thus gives a lower bound for the actual SN of the state.
As a general trend, we find an increasing SN if the polariton splitting to binding energy ratio $p_s$ tends to zero.
Moreover, a negative detuning makes the entanglement more robust against dephasing, at least if $p_s \leq 2$.
This means that the initial state of the branch-entangled polariton pairs is given by a superposition of maximally entangled Bell states if $\delta < -1$ and $p_s \to 0$.

\section{\label{sec:comb}Multipartite entanglement with frequency comb pulses}
\begin{figure}
  \includegraphics[width=\linewidth]{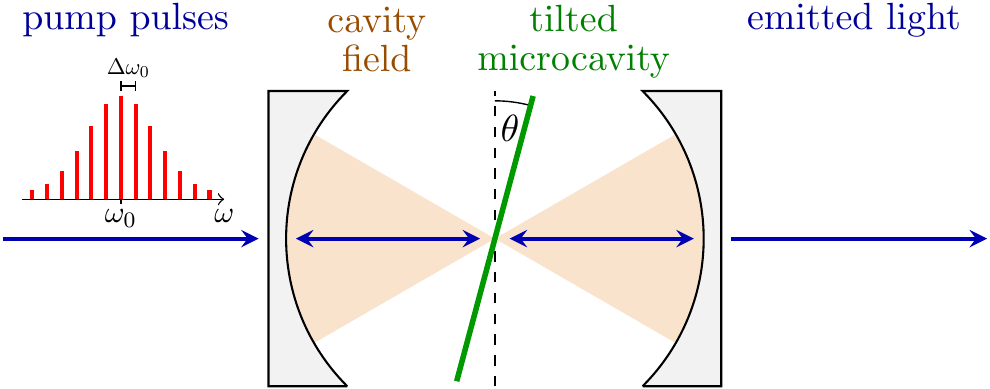}
  \caption{\label{fig:FreqCombSetup}Sketch of the setup considered for generation of an entangled frequency comb.}
\end{figure}

We now extend the system under consideration by placing the planar microcavity inside a cavity and tilting it by an angle $\theta$ from the vertical related to the incident light (see Fig.~\ref{fig:FreqCombSetup}).
The pump pulse itself represents a frequency comb with central frequency $\omega_0$ and line separation $\Delta\omega_0$.

\subsection{\label{ssec:phasematch}Polariton interaction}
\begin{figure}
  \includegraphics[width=\linewidth]{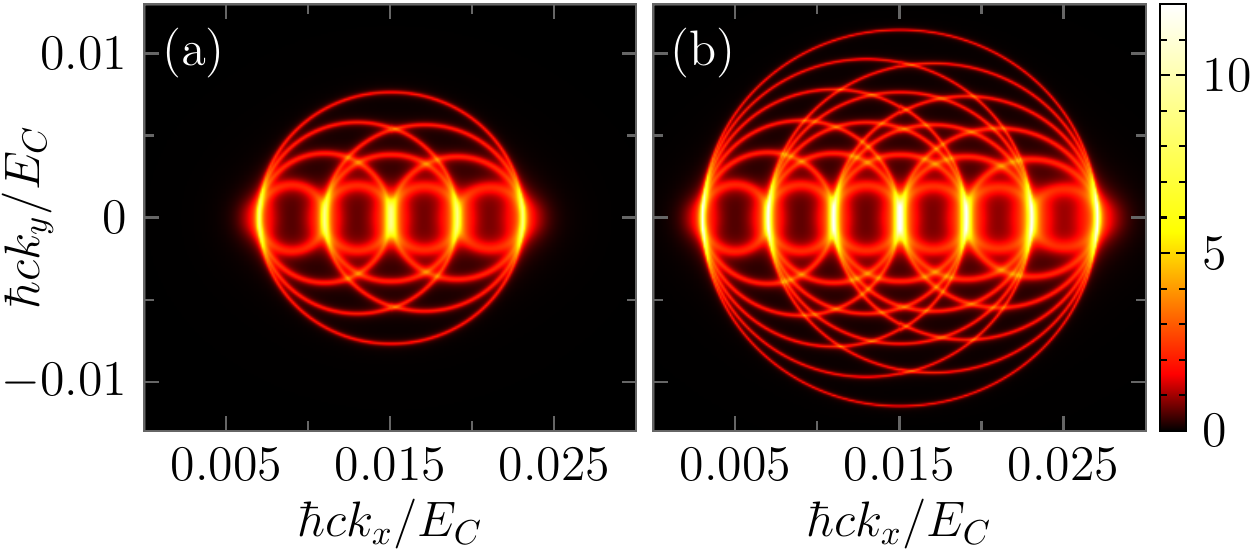}
  \caption{\label{fig:phasematch}Phase matching in dependence on the in-plane wave-vector $\mathbf{k} = (k_x, k_y)^T$ for (a) $N = 2$ and (b) $N = 3$ modes. 
    System parameters are $E_C = E_X = 1.5\,$eV, $\hbar \Omega = 2\,$meV, $\gamma = 1\,\upmu$eV, $\hbar c k_0 / E_C = 0.015$, and $\hbar c \Delta k_0 / E_C = 0.004$.}
\end{figure}

Defining an in-plane wave-vector $k_0 = (\omega_0 / c) \sin \theta$ and $\Delta k_0 = (\Delta\omega_0 / c) \sin\theta$ we find that pumping the microcavity with a frequency comb generates polaritons at the equidistant points $k_n \equiv k_{\text{p}n} = k_0 + n \Delta k_0$ in $k$-space, where $n \in [-N, N]$ and $N$ is the number of (pump) spectral lines above and below the central frequency $\omega_0$.
Phase-matching scattering processes between all frequencies of the comb can be achieved when the lower polariton branch is pumped below the magic angle.
The scattering channels in line with that are depicted in Fig.~\ref{fig:phasematch}, which visualizes the phase-matching function
\begin{multline}\label{eta}
  \eta(\mathbf{k}) = \sum_{m,n=-N}^N \gamma^2 \Big\{ \big[ E_1(\mathbf{k}) + E_1(\mathbf{k}_n + \mathbf{k}_m - \mathbf{k}) \\
    - E_1(\mathbf{k}_n) - E_1(\mathbf{k}_m) \big]^2 + \gamma^2 \Big\}^{-1}.
\end{multline}
In Eq.~\eqref{eta}, $\gamma$ denotes the polariton broadening.

In what follows, we consider the case $k_y = 0$ for $N = 2$ in Fig.~\ref{fig:phasematch}(a) and $N = 3$ in Fig.~\ref{fig:phasematch}(b).
We expect, that the pumped modes become entangled because the different mixed-pump scattering processes are indistinguishable.
Then, the Hamiltonian describing the polaritons and their scattering processes takes the form
\begin{eqnarray}\label{H}
  \hat{H} &=& \sum_{n = -N}^N E_1(\mathbf{k}_n) \Big( \hat{a}_n^\dagger \hat{a}_n^{} + \frac{1}{2} \Big) \nonumber\\
    && + \sum_{\substack{m, n = -N \\ m \neq n}}^N \big( R_{m,n}^{} \hat{a}_m^\dagger \hat{a}_n^\dagger + \text{H.\,c.} \big) \,,
\end{eqnarray}
where $\hat{a}_n^\dagger \equiv \hat{a}_{1, \text{p}n}^\dagger$ ($\hat{a}_n \equiv \hat{a}_{1, \text{p}n}$) creates (annihilates) a polariton in the lower branch with wave-vector $\mathbf{k}_n = (k_0 + n \Delta k_0) \mathbf{e}_x = k_n \mathbf{e}_x$.
The coupling of the different modes is given by the (real) matrix elements
\begin{equation}
  R_{m,n} = \frac{1}{2} \frac{R_X^2}{A} V_{\mathbf{k}_m, \mathbf{k}_n, \mathbf{k}_n-\mathbf{k}_m}^{1 1 1 1} P_m P_n = R_{n,m} \,.
\end{equation}
Here, $R_X$ is the radius of the semiconductor exciton, $A$ is the sample surface, and $P_n$ is the pump spectral amplitude of mode $n$.

\subsection{\label{ssec:groundstate}Polariton ground state}
The ground state of the Hamiltonian\ \eqref{H} is a Gaussian state.
Introducing the amplitude $\hat{q}_n = \sqrt{\hbar / 2} (\hat{a}_n^{} + \hat{a}_n^\dagger)$ and the phase quadrature $\hat{p}_n = -\rmi \sqrt{\hbar / 2} (\hat{a}_n^{} - \hat{a}_n^\dagger)$ of the individual modes, we define a vector of quadratures $\hat{\boldsymbol{\xi}} = (\hat{q}_{-N}, \dots, \hat{q}_N, \hat{p}_{-N}, \dots, \hat{p}_N)^T$.
Its expectation values
\begin{equation}
  \Sigma_{i,j} = \frac{1}{2} \big\langle \hat{\xi}_i \hat{\xi}_j + \hat{\xi}_j \hat{\xi}_i \big\rangle - \big\langle \hat{\xi}_i \big\rangle \big\langle \hat{\xi}_j \big\rangle \,,
\end{equation}
are the entries of the Gaussian state's covariance matrix $\boldsymbol{\Sigma} = \big\langle \hat{\boldsymbol{\xi}} \hat{\boldsymbol{\xi}}^T \big\rangle - \big\langle \hat{\boldsymbol{\xi}} \big\rangle \big\langle \hat{\boldsymbol{\xi}} \big\rangle^T$, where $(\hat{\boldsymbol{\xi}} \hat{\boldsymbol{\xi}}^T)_{i,j} = (\hat{\xi}_i \hat{\xi}_j + \hat{\xi}_j \hat{\xi}_i) / 2$.
To determine the covariance matrix, we rewrite the Hamiltonian\ \eqref{H} using quadrature operators,
\begin{eqnarray}
  \hat{H} &=& \sum_{n = -N}^N \frac{E_1(k_n)}{2 \hbar} \big( \hat{q}_n^2 + \hat{p}_n^2 \big) \nonumber\\
    && + \sum_{\substack{m, n = -N \\ m \neq n}}^N \frac{\Re R_{m,n}}{\hbar} (\hat{q}_m \hat{q}_n - \hat{p}_m \hat{p}_n) \nonumber\\
    &=& \hat{\boldsymbol{\xi}} \cdot \mathbf{H} \hat{\boldsymbol{\xi}} \,,
\end{eqnarray}
where the real symmetric ($\mathbf{H} = \mathbf{H}^T$) and positive definite ($\mathbf{H} > 0$) matrix $\mathbf{H}$ has a block structure
\begin{equation}
  \mathbf{H} = \frac{1}{\hbar} \begin{pmatrix} \mathbf{H}_{q,q} & \mathbf{0} \\ \mathbf{0} & \mathbf{H}_{p,p} \end{pmatrix} \,.
\end{equation}
The ground state is given by the normalized, Gaussian wave-function
\begin{equation}\label{psi}
  \psi(\mathbf{q}) = \sqrt[4]{\frac{\det\mathbf{V}}{\pi^{2N+1}}} \exp \Big( -\frac{1}{2} \mathbf{q} \cdot \mathbf{V} \mathbf{q} \Big) \,,
\end{equation}
where $\mathbf{q} = (q_{-N}, \dots, q_N)$ are the coordinates, and $\mathbf{V}$ is also a real symmetric and positive definite matrix.
Then the corresponding eigenvalue problem reads
\begin{eqnarray}
  \hat{H} \psi(\mathbf{q}) &=& \frac{1}{\hbar} \Big[ \mathbf{q} \cdot \mathbf{H}_{q,q} \mathbf{q} - \hbar^2 \mathbf{q} \cdot \mathbf{V} \mathbf{H}_{p,p} \mathbf{V} \mathbf{q} \nonumber\\
    && + \hbar^2 \text{Tr} \big( \mathbf{H}_{p,p} \mathbf{V} \big) \Big] \psi(\mathbf{q}) \nonumber\\
    &=& E \psi(\mathbf{q}) \,.
\end{eqnarray}
This equation is solved by the ground state $\psi(\mathbf{q})$ if
\begin{equation}
  \mathbf{V} = \frac{1}{\hbar} \mathbf{H}_{p,p}^{-1/4} \mathbf{H}_{q,q}^{1/2} \mathbf{H}_{p,p}^{-1/4} \,.
\end{equation}
The corresponding ground-state energy is
\begin{equation}
  E = \text{Tr} \big( \sqrt{\mathbf{H}_{q,q} \mathbf{H}_{p,p}} \big) \,.
\end{equation}
Finally, the covariance matrix $\boldsymbol{\Sigma}$ follows from transformation to the Wigner representation:
\begin{equation}
  \boldsymbol{\Sigma} = \frac{1}{2} \begin{pmatrix} \mathbf{V}^{-1} & 0 \\ 0 & \hbar^2 \mathbf{V} \end{pmatrix} \,.
\end{equation}

\subsection{\label{ssec:entanglement}Multipartite entanglement between continuous variables}
The complexity of entanglement is realized if a system consisting of a manifold of parties is considered.
For example, a subsystem can be entangled with a second one but separated from a third one.
However, the third subsystem may still be entangled with the second one.
The number of such different forms of multipartite entanglement rapidly increases with the number of parties.
Nevertheless, different techniques to tackle entanglement in such systems have emerged~\cite{asi04, gsvcrtf15, sv13}.

To determine the multipartite continuous-variable entanglement of the polariton ground state~\eqref{psi} we again use the witnessing method~\cite{sv13, gsvcrtf15}.
In particular, the Gaussian state $\hat{\rho}$ with covariance matrix $\boldsymbol{\Sigma}$ is entangled with respect to a $K$ partition if and only if a Hermitian operator $\hat{L}$ with
\begin{equation}\label{ent-cond}
  \text{Tr} \big( \hat{L} \hat{\rho} \big) < g_{\mathcal{I}_1:\dots:\mathcal{I}_K}^\text{min}
\end{equation}
exists~\cite{hhh96, hhh01}.
In Eq.~\eqref{ent-cond}, the $2 N + 1$ modes are partitioned into $K \in [1, 2 N + 1]$ different and complementary subsystems $\mathcal{I}_1:\dots:\mathcal{I}_K$ and $g_{\mathcal{I}_1:\dots:\mathcal{I}_K}^\text{min}$ is the minimum expectation value of $\hat{L}$ among all separable states of this partition.
Because the  first-order moments are irrelevant for entanglement,  without loss of generality, we may assume a state $\hat{\rho}$ with
\begin{equation}\label{ent-2}
  \text{Tr}(\hat{\rho} \hat{\boldsymbol{\xi}}) = 0.
\end{equation}

Next, we need to find a particular Hermitian operator $\hat{L}$ that fulfills Eq.~\eqref{ent-cond}.
According to Ref.~\cite{sv13}, the optimal choice is $\hat{L} = \hat{H}$ for the detection of entanglement within the ground state~\eqref{psi} of $\hat{H}$.
Then, the left-hand side of condition~\eqref{ent-cond} readily follows as
\begin{equation}\label{L}
  \langle \hat{L} \rangle = \text{Tr} \big( \hat{L} \hat{\rho} \big) = \text{Tr} (\mathbf{H} \boldsymbol{\Sigma}) \,,
\end{equation}
and the right-hand side becomes~\cite{sv13, gsvcrtf15}
\begin{equation}\label{g}
  g_{\mathcal{I}_1:\cdots:\mathcal{I}_K}^\text{min} = \sum_{j=1}^K \text{Tr} \bigg( \sqrt{\mathbf{H}_{q,q}^{\mathcal{I}_j} \mathbf{H}_{p,p}^{\mathcal{I}_j}} \bigg) \,.
\end{equation}
Here, the submatrix of $\mathbf{A}$ that contains only the rows and columns belonging to the index set $\mathcal{I}_j$ is $\mathbf{A}^{\mathcal{I}_j}$.

With Eqs.~\eqref{ent-cond},~\eqref{L}, and~\eqref{g} we have all ingredients for the detection of multipartite entanglement.
To visualize the results we introduce the so-called entanglement visibility~\cite{sw16}
\begin{equation}
  \mathcal{V} = \frac{g_{\mathcal{I}_1:\cdots:\mathcal{I}_K}^\text{min} - \langle \hat{L} \rangle}{g_{\mathcal{I}_1:\cdots:\mathcal{I}_K}^\text{min} + \langle \hat{L} \rangle} \,,
\end{equation}
that quantifies the fulfillment of condition~\eqref{ent-cond}.
$\mathcal{V}$ takes positive (negative) values if the state $\hat{\rho}$ is entangled (not entangled) with respect to the chosen $K$ partition.
A large positive value of $\mathcal{V}$ indicates a sufficient separation between the left-hand and right-hand sides of Eq.~\eqref{ent-cond}.
The closer $\mathcal{V}$ is to zero the higher is the required resolution of a corresponging detector that measures $\langle \hat{L} \rangle$.
Nevertheless, $\mathcal{V}$ does not quantify multipartite entanglement.
To the best of our knowledge, a single universal measure of multipartite entanglement does not exist so far.

\begin{figure}
  \includegraphics[width=0.8\linewidth]{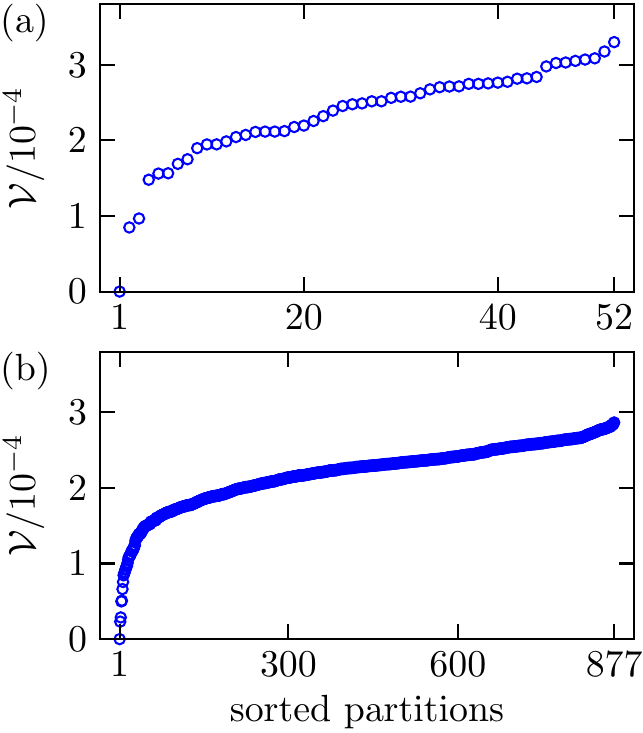}
  \caption{\label{fig:entanglement}Visibility of each partition's entanglement for (a) $N = 2$ and (b) $N = 3$ yielding a total of (a) 52 and (b) 877 possible partitions.
    System parameters are the same as in Fig.~\ref{fig:phasematch}.}
\end{figure}

In Fig.~\ref{fig:entanglement} we plot the visibility of each partition's entanglement for $N = 2$ and $N = 3$ with a total of $2 N + 1 = 5$ and $7$ modes, respectively.
The so-called Bell number of possible partitions is determined from combinatorics.
For example, the set $\{1,2,3\}$ of three modes has five partitions: 1:2:3, 12:3, 13:2, 1:23, and 123.
Likewise, one obtains 52 (877) partitions for a set of five (seven) modes.

Since $\mathcal{V} \geq 0$ for each partition in Fig.~\ref{fig:entanglement}, we can conclude that each subsystem is entangled with all the other subsystems.
In Figs.~\ref{fig:entanglement}(a) and \ref{fig:entanglement}(b), the points with $\mathcal{V} = 0$  correspond to the trivial partition $K = 1$, where all modes are incorporated in a single party only.
This partition is necessarily not entangled.
All other partitions have a positive entanglement visibility, such that the polariton ground state is proven to be fully multipartite entangled.
In other words, entanglement will be distributed among each mode of the initial frequency comb pump pulse once the light interacts with the microcavity.

\section{\label{sec:dyn}Dissipative buildup of entanglement in a Dicke-type cavity}
\begin{figure}
  \includegraphics[width=\linewidth]{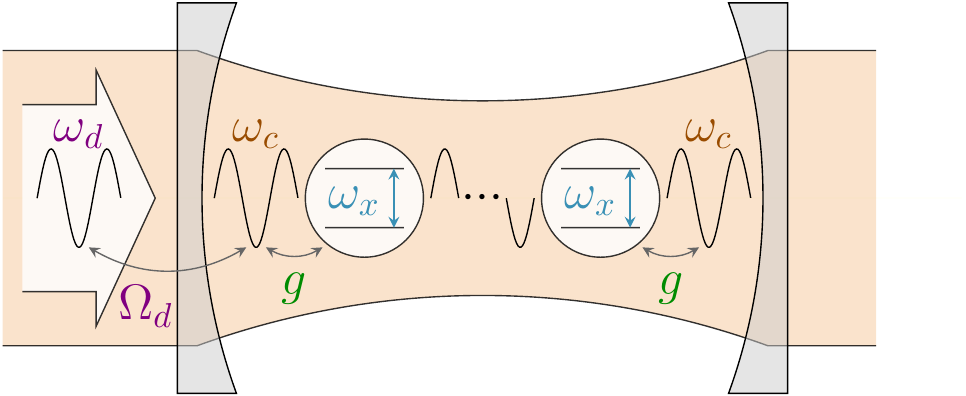}
  \caption{\label{fig:cavity}Two-level emitters in a laser-driven cavity.}
\end{figure}

The results presented so far do not take account of effects arising from dissipation, that is, from the coupling to environmental degrees of freedom.
Therefore, in this section, we study the dissipative dynamics of entanglement in the strong light-matter coupling regime.
In particular, we are interested in the entanglement between two two-level emitters (qubits) that are placed inside a cavity which itself is driven by an external laser, see Fig.~\ref{fig:cavity}.
Such a situation can be described by the driven Dicke model~\cite{dic54}:
\begin{eqnarray}\label{HD}
  \hat{H}_D(t) &=& \hbar \omega_c \hat{c}^\dagger \hat{c} + \hbar \omega_x \sum_{j=1}^2 \hat{\sigma}_+^{(j)} \hat{\sigma}_-^{(j)} \nonumber\\
    && + \hbar g (\hat{c} + \hat{c}^\dagger) \sum_{j=1}^2 \big( \hat{\sigma}_-^{(j)} + \hat{\sigma}_+^{(j)} \big) \nonumber\\
    && + \hbar \Omega_d (\hat{c} + \hat{c}^\dagger) \cos \omega_d t \,.
\end{eqnarray}
In Eq.~\eqref{HD}, $\hat{c}^\dagger$ ($\hat{c}$) creates (annihilates) a cavity photon with frequency $\omega_c$, $\hat{\sigma}_-^{(j)}$ ($\hat{\sigma}_+^{(j)}$) incorporates the relaxation (excitation) of the $j$th emitter with transition energy $\omega_x$.
The emitter-cavity coupling strength is $g$.
The driving amplitude and frequency are denoted by $\Omega_d$ and $\omega_d$, respectively.
In what follows we only consider the resonance situation: $\omega_r = \omega_c = \omega_x = \omega_d$.

A consistent theoretical description of strongly interacting quantum optical systems coupled to the environment is difficult, even in the simplest case,  when dissipation can be treated in the Markovian approximation (for previous approaches, see, e.\,g., Refs.~\cite{car99, bp02, wei12, aga13, sch14}).
Starting from the Dicke model~\eqref{HD} without drive $\Omega_d = 0$, it is necessary to replace the standard quantum optical master equation by a master equation expressed in the photon-dressed emitter states~\cite{bgb11, alz06, sb08, rsh13}.
Combining this equation with the input-output formalism~\cite{cg84, gc85, gra89, sg96pra}, the statistics of the emitted photons can be evaluated, showing that cooperative effects lead to the generation of nonclassical light already at weak light-matter coupling if the number of emitters is increased~\cite{paf15}.
For finite driving $\Omega_d \neq 0$ the approach developed in Ref.~\cite{paf17} can be used to monitor  the dynamics of the system, e.g., in order to analyze the dynamic Stark effect via the  shifting of spectral lines in the emission spectrum.

Here, we consider a Dicke-type cavity coupled to bosonic environmental degrees of freedom with interaction Hamiltonian $H_I = -\rmi \hat{X} \sum_\alpha \lambda_\alpha (\hat{b}_\alpha - \hat{b}_\alpha^\dagger)$ where $\hat{X}$ is a Hermitian system operator and $\hat{b}_\alpha$ are annihilation operators of environment photons at frequencies $\omega_\alpha$ with coupling constants $\lambda_\alpha$.
We choose $\hat{X} = -\rmi (\hat{c} - \hat{c}^\dagger)$ for the interaction of the cavity and $\hat{X} = -\rmi (\hat{\sigma}_-^{(j)} - \hat{\sigma}_+^{(j)})$ for the interaction of the $j$th emitter with the environment.

For weak system-environment coupling, the dissipative dynamics of the system can be described by a Markovian master equation.
Since we have assumed a periodic time-dependence $\hat{H}_D(t) = \hat{H}_D(t + T_d)$ with period $T_d = 2 \pi / \omega_d$ in Eq.~\eqref{HD}, the Floquet states,
\begin{equation}
  | \psi_n(t) \rangle = \rme^{-\rmi \epsilon_n t} | \phi_n(t) \rangle
\end{equation}
with quasienergies $\epsilon_n \in \mathbb{R}$ and periodic wavefunctions $| \phi_n (t) \rangle = | \phi_n(t + T_d) \rangle$, can be used as computational basis.
The master equation then decouples into the two equations of motion~\cite{paf17}
\begin{subequations}\label{master}\begin{eqnarray}\label{master-nn}
  \frac{\rmd}{\rmd t} \rho_{n, n}(t) &=& \sum_{k, \nu} \chi(\omega_{k, n, \nu}) |X_{n, k, \nu}|^2 \rho_{k, k}(t) \nonumber\\
    && - \sum_{k, \nu} \chi(\omega_{n, k, \nu}) |X_{k, n, \nu}|^2 \rho_{n, n}(t) \,, \\\label{master-mn}
  \frac{\rmd}{\rmd t} \rho_{m, n}(t) &=& -Z_{m, n} \rho_{m, n}(t) \qquad (m \neq n)
\end{eqnarray}\end{subequations}
for the density matrix elements $\rho_{m, n}(t) = \langle \psi_m(t) | \hat{\rho}(t) | \psi_n(t) \rangle$.
In Eqs.~\eqref{master}, we introduced the transition energy $\omega_{m, n, \nu} = \epsilon_m - \epsilon_n + \nu \omega_d$ where $\nu$ counts the Fourier modes of the periodic states
\begin{equation}
  | \phi_n(t) \rangle = \sum_{\nu=-\infty}^\infty \rme^{-\rmi \nu \omega_d t} | \tilde{\phi}_n(\nu) \rangle \,.
\end{equation}
Accordingly, the transition matrix elements are
\begin{equation}
  X_{m, n, \nu} = \sum_\mu \langle \tilde{\phi}_m(\mu - \nu) | X | \tilde{\phi}_n(\mu) \rangle \,,
\end{equation}
and the coefficients of the non-diagonal equations of motion~\eqref{master-mn} read
\begin{eqnarray}
  Z_{m, n} &=& \frac{1}{2} \sum_{k, \nu} \big[ \chi(\omega_{m, k, \nu}) + \rmi \xi(\omega_{m, k, \nu}) \big] |X_{k, m, \nu}|^2 \nonumber\\
    && + \frac{1}{2} \sum_{k, \nu} \big[ \chi(\omega_{n, k, \nu}) + \rmi \xi(\omega_{n, k, \nu}) \big] |X_{k, n, \nu}|^2 \nonumber\\
    && - \sum_\nu \chi(\nu \omega_d) X_{m, m, \nu} X_{n, n, \nu}^* \,.
\end{eqnarray}
The functions $\chi(\omega)$ and $\xi(\omega)$ are even and odd Fourier transforms of the reservoir correlation function for a thermal environment at temperature $T$, and are given by
\begin{equation}
  \chi(\omega) = \begin{cases} \gamma(\omega) [n(\omega, T) + 1] & \text{if}~\omega > 0 \\ \gamma(-\omega) n(-\omega, T) & \text{if}~\omega < 0 \end{cases}
\end{equation}
and
\begin{equation}
  \xi(\omega) = \begin{cases} \Re \Gamma(\omega + \rmi 0^+) [n(\omega, T) + 1] & \text{if}~\omega > 0 \\ \Re \Gamma(-\omega + \rmi 0^+) n(-\omega, T) & \text{if}~\omega < 0 \end{cases} \,,
\end{equation}
respectively.
In these equations, $\gamma(\omega)$ is the environment spectral function, $\Gamma(\omega)$ is its analytic continuation into the upper half plane, and $n(\omega,T) = (\rme^{\omega/T} - 1)^{-1}$ is the thermal distribution function.
Here, we assume the same Ohmic spectral function $\gamma(\omega) = \lambda \omega / \omega_0$ for both cavity and emitter environments.
The respective environment temperatures are also identical.

\begin{figure}
  \includegraphics[width=0.9\linewidth]{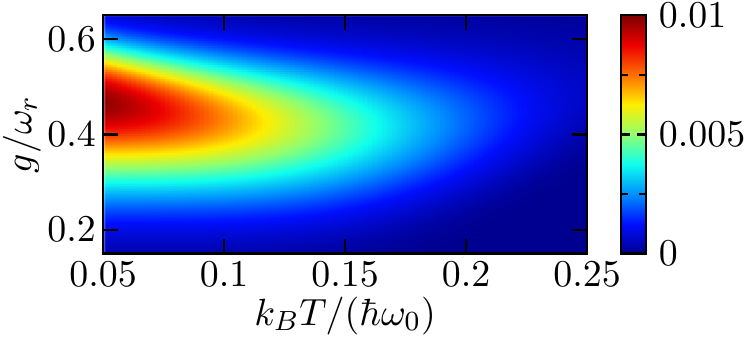}
  \caption{\label{fig:EOF}EOF of two emitters (for $t \to \infty$) as a function of the environment temperature $T$ and the emitter-cavity coupling strength $g$.
    The laser intensity $\Omega_d = 10^{-4}\,\omega_r$.}
\end{figure}

\begin{figure}
  \includegraphics[width=0.9\linewidth]{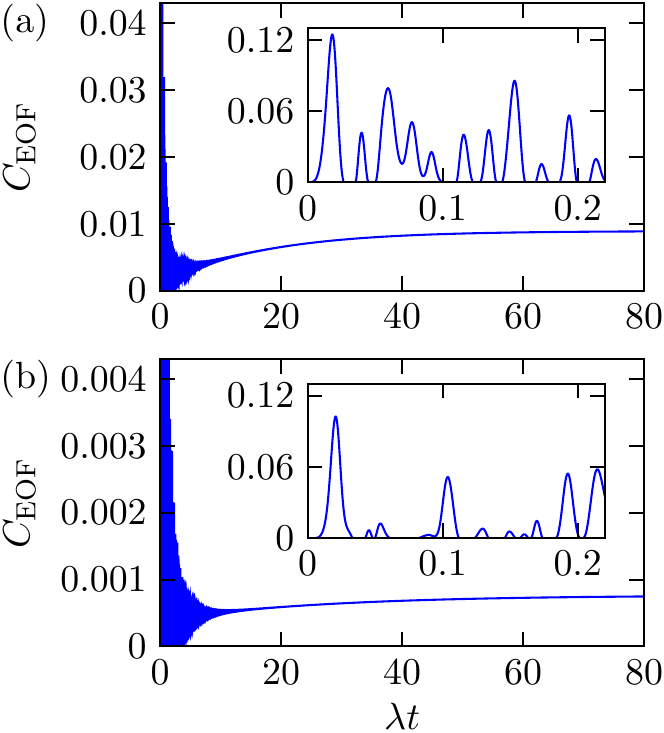}
  \caption{\label{fig:ent_dyn}Time-dependence of the EOF for two emitters, where (a) $g = 0.45 \, \omega_r$, $k_B T = 0.07 \, \hbar \omega_r$ and (b) $g = 0.55 \, \omega_r$, $k_B T = 0.22 \, \hbar \omega_r$.
    Other model parameters are $\Omega_d = 10^{-4} \, \omega_r$ and $\lambda = 10^{-2} \, \omega_r$.}
\end{figure}

The decoupling of diagonal and non-diagonal density matrix elements allows for a straightforward determination of the system state in the long-time limit ($t \to \infty$).
To quantify the bipartite entanglement between the two emitters we evaluate the partial trace over the cavity modes and calculate the entanglement of formation (EOF)~\cite{woo98, nc10},
\begin{equation}
  C_\text{EOF} = -\eta \log_2 \eta - (1 - \eta) \log_2 (1 - \eta) 
\end{equation}
with $\eta = (1 + \sqrt{1 - C^2}) / 2$, as well as the concurrence~\cite{hw97}
\begin{equation}
  C = \max\{ 0, \lambda_1 - \lambda_2 - \lambda_3 - \lambda_4\} \,.
\end{equation}
Here, $\lambda_i$ for $i = 1, \dots, 4$ are the eigenvalues, in decreasing order, of the Hermitian matrix
\begin{equation}
  \sqrt{\sqrt{\hat{\rho}_x} (\hat{\sigma}_y \otimes \hat{\sigma}_y) \hat{\rho}_x^* (\hat{\sigma}_y \otimes \hat{\sigma}_y) \sqrt{\hat{\rho}_x}} \,,
\end{equation}
where $\hat{\sigma}_y$ is a Pauli matrix and $\hat{\rho}_x$ is the density operator of the two emitters.
Figure~\ref{fig:EOF} gives the EOF in dependence on the environment temperature $T$ and the emitter-cavity coupling strength $g$.
While a considerable amount of entanglement is found at strong coupling and low temperature, almost no entanglement is generated at ultrastrong couplings.
This result seems to be in contradiction with a recent experimental finding where the entanglement between a single emitter and the cavity mode becomes large at ultrastrong coupling~\cite{yfakss17}.
Nevertheless, entanglement between two emitters, which is obtained after tracing out the cavity degrees of freedom, is a different topic.
For example, in Ref.~\cite{ash14} it was found that the emitter's excitation probability starts to decrease above a certain coupling strength.
Although entanglement additionally involves emitter correlations, the statement in Ref.~\cite{ash14} gives a first hint why $C_{\text{EOF}}$ goes to zero for large $g$.
The real reason is, that for the low laser intensities $\Omega_d \ll \omega_r, g$ used to produce Fig.~\ref{fig:ent_dyn} the asymptotic state is almost thermal and the eigenstates of the Dicke model without laser drive do not include emitter entanglement at ultrastrong emitter-cavity coupling.
Note that a completely different behavior will show up in the time-evolution of the EOF, where the transitions between Floquet states become important.

The dynamics of the EOF is obtained from the solution of the Markovian master equation~\eqref{master}.
The system is prepared in a product state, where the cavity modes are thermally distributed at the environment temperature $T$ and the two-level emitters are in their ground states.
The resulting time-dependence of the EOF of the two emitters is shown in Figs.~\ref{fig:ent_dyn}(a) and \ref{fig:ent_dyn}(b).
In the short time dynamics, the EOF shows a non-monotonic, oscillating behavior indicating the relevance of the transitions between different Fourier modes.
Nevertheless, the EOF stays positive for all times.
The stationary value for $t \to \infty$ coincides with the results depicted in Fig.~\ref{fig:EOF}.
Importantly, the largest EOF is obtained within the first few oscillation periods.
Although the amount of entanglement in the asymptotic state is pretty small and perhaps might not be detectable in experiments, the entanglement created in the short-time evolution of the system is at least one order of magnitude higher.
Furthermore, the increased generation of entanglement within the first few oscillation periods provides a route towards optimizing the laser-induced emitter entanglement using short pump pulses and thereby successively enhances the EOF.

\section{\label{sec:conc}Conclusions}
We have discussed the generation and characterization of entangled light from driven planar semiconductor microcavities.
Pumping of the microcavity by a pulse train leads to a simultaneous creation of multiple branch entangled polariton pairs when phase-matching interaction processes are at play.
The coupling of the intra-cavity scattering dynamics to an external field then transfers these quantum correlations to frequency-entangled photons that can be easily detected.
The generation of multiple pairs of entangled polaritons permits the simultaneous creation of copies of entangled qubit states, which is a basic requirement of quantum information processing.

Placing the planar microcavity inside an optical cavity designed such that a number of copropagating frequency modes is resonant, the phase-matching scattering processes can be tuned to entangle every mode of the frequency comb.
In contrast to setups employing parametric down-conversion in nonlinear crystals, our cavity quantum-electrodynamic based approach will not change the frequencies of the incident pump pulses.
It therefore yields an increased photon efficiency that might be exploited in quantum-communication experiments.

The non-classical correlations of the polariton pairs can be detected by a Schmidt decomposition.
We found, that the Schmidt number approaches a maximum if the detuning of exciton and photon energies is negative and if the ratio of polariton splitting to exciton binding energy tends to zero.
Multipartite entanglement in the continuous variable systems under study was identified via the separability eigenvalue equations and specified by the entanglement visibility.
We used this approach to proof the entangling of all modes of a frequency comb.
Clearly the entanglement visibility should be optimized by tuning the model parameters of the proposed quantum-well-cavity setup in future experiments.
Theoretically, we performed such an optimization for a microcavity driven by a pump pulse train.

Since actual experiments always suffer from losses, we studied the impact of dissipation on the generation of entanglement.
To calculate the entanglement dynamics in the weak system-environment coupling regime, we consider a time-periodic perturbation (driving) and make use of the Floquet master equation.
For the paradigmatic driven Dicke cavity system we found, that a large amount of entanglement is created within the first few laser oscillation periods before it decays oscillatory towards its stationary value in the long-time limit.
Having this in mind it would be instructive to consider a sequence of pump pulses in order to successively increase the produced entanglement.
To capture such a scenario, the Floquet description of a perfectly periodic Hamiltonian may be combined with an adiabatic treatment of the changing driving amplitude.

\begin{acknowledgments}
This work was granted by Deutsche Forschungsgemeinschaft through SFB 652 (project B5).
D.\ P.\ is also grateful for support by SFB/TRR 24 (project B10).
\end{acknowledgments}


%

\end{document}